# Thin-film based ultralow noise SQUID magnetometer

M. Schmelz, V. Zakosarenko, A. Chwala, T. Schönau, R. Stolz, S. Anders, S. Linzen and H.-G. Meyer

*Abstract*— We report on the development of an ultralow noise thin-film based SQUID magnetometer. A niobium thin-film pickup coil is connected to the input coil of a SQUID current sensor. The low capacitance of the used sub-micrometer cross-type Josephson junctions enable superior noise performance of the device. Application scenarios e.g. in geophysics and ultra-low field magnetic resonance imaging are discussed.

## I. INTRODUCTION

SUPERCONDUCTING quantum interference devices (SQUIDs) are known to be today's most sensitive devices for the detection of magnetic flux $\Phi$. They convert magnetic flux or any physical property that can be transformed into magnetic flux, for example magnetic flux density $B$, into e.g. a voltage across the device.

The figure of merit for a magnetometer is the magnetic field noise spectral density $S_B$, which can be expressed as $S_B^{1/2} = S_\Phi^{1/2}/A_{eff}$. Here $A_{eff} = \Phi/B$ being the effective area of the SQUID magnetometer [1]. Modern SQUID magnetometer systems already exhibit magnetic field noise level down to 1 fT/Hz$^{1/2}$ [2]. They are used e.g. in geophysical prospection, biomagnetism or ultra-low field magnetic resonance imaging (ULF MRI).

To meet the requirements for future demanding applications, during past years, there have been continuous efforts to improve the magnetic field noise $S_B^{1/2}$ of SQUID magnetometers. The improved sensitivity of such SQUID systems may enable them to be used in passive electromagnetic methods, such as audiofrequency magnetics or magnetotellurics, which would benefit from the SQUID performance especially at low frequencies, resulting in larger depth of investigation.

According to [3], beside the cooling-issue today's biggest challenge for bringing ULF MRI into clinical practice is the enhancement of the signal-to-noise-ratio (*SNR*), which determines the imaging time $t \propto 1/(SNR)^2$. In order to increase the $SNR \propto B_p/B_S$, one needs to lower the SQUID system noise $B_S = S_B^{1/2}$ and to increase the strength of the pre-polarizing field $B_p$. Acceptable imaging times for ULF MRI demand for $B_S$ of the order of 0.1 fT/Hz$^{1/2}$, comparable to the limit set by the body noise [3].

While increasing the pre-polarizing fields up to the order of 100 mT, most of the currently used SQUIDs utilizing wire-wound type-II superconducting pickup coils show degraded low-frequency noise performance due to flux trapping in the pickup loop [4]. Although the recently investigated type-I superconducting pickup coils made of lead do not show a degraded performance like their type-II counterparts [5], the lack of a suitable wire technology and their fragile nature make them less attractive.

Replacing them with thin-film versions of the pickup loop enable common provisions against flux trapping, such as e.g. narrow linewidth structures of the superconductor [6-8].

Within this paper we will therefore focus on a SQUID magnetometer composed of a SQUID current sensor and a planar niobium thin-film pickup loop. Superior noise performance is achieved due to the implementation of low capacitance cross-type Josephson junctions. In section II we will introduce the sensor design. Measurement results on the fabricated SQUID magnetometer demonstrating an exceptionally low magnetic field noise will be given in section III.

## II. SENSOR DESIGN

The magnetic field noise spectral density $S_B$ can be improved in two ways: (i) increase of the effective area of the SQUID magnetometer and (ii) by improving the SQUID's noise performance.

To avoid very large pickup-loops, which may impede e.g. the integration of an orthogonal triple of such sensors into a SQUID system aimed e.g. for geophysical measurements, within this paper we will mainly focus on the noise improvement of the used SQUIDs. The presented magnetometer is composed of a SQUID current sensor and a planar niobium thin-film pickup loop implemented on two separate chips. For application in ULF MRI these planar pickup loop may be replaced e.g by an axial-gradiometric thin-film version as for example presented in [9].

Fig. 1 shows two photographs of the investigated device. The SQUID current sensor is located on the small chip to the left, with dimensions of 2.5 mm × 2.5 mm. The thin-film pickup coil, located on the right hand side silicon chip, has outer dimensions of 29.5 mm × 33.5 mm. The pickup coil is connected via superconducting Nb bond wires to the on-chip

M. Schmelz, A. Chwala, T. Schönau, R. Stolz, S. Anders, S. Linzen and H.-G. Meyer are with Leibniz Institute of Photonic Technology, Jena, Germany.
V. Zakosarenko is with the Supracon AG, Jena, Germany.
The corresponding author: M. Schmelz (phone: +49 3641 206 122; fax: +49 3641 206 199; e-mail: matthias.schmelz@ipht-jena.de).

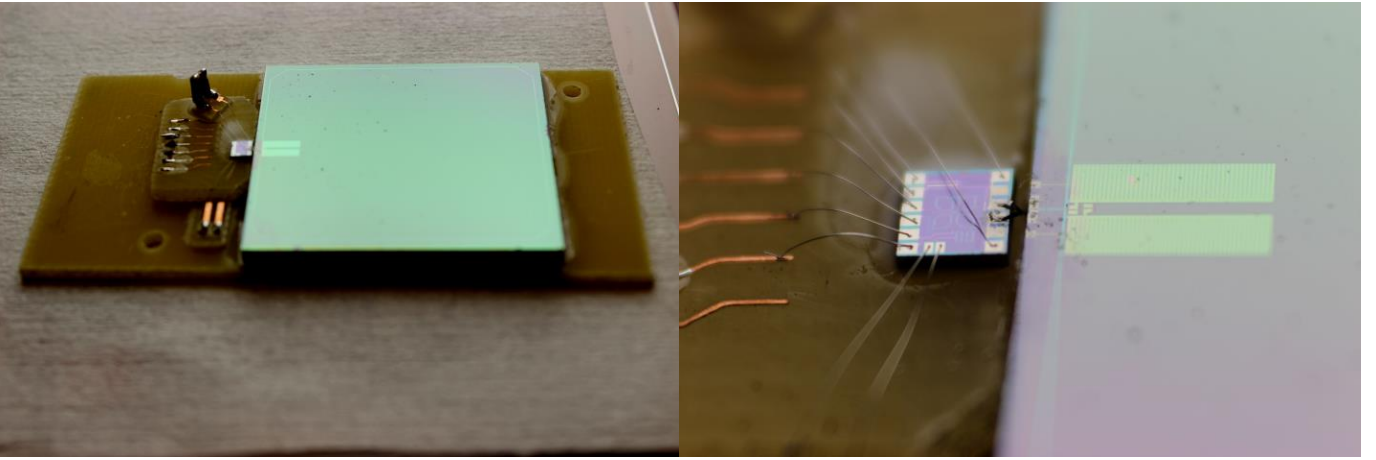

Fig. 1. The left photograph shows the SQUID magnetometer composed of a SQUID current sensor and a thin-film pickup loop presented by the left small and right chip, respectively. Both chips are connected via superconducting Nb wire bonds. Right photograph shows a close-up of the SQUID current sensor chip on the left and the blueish thin-film pickup loop with integrated RC-filter on the right hand side.

input coil on top of the SQUID washer. Both chips are glued on a chip carrier of glass-fiber reinforced plastic. The right photograph shows a close-up of the current sensor SQUID. On the right hand side one can moreover see the blueish thin-film pickup loop with integrated RC-filter used to obtain smooth flux-voltage characteristics of the SQUID. To avoid too large screening currents in the input coil of the SQUID current sensor, the SQUID chip feature an integrated resistive heater next to the SQUID input coil.

The SQUID current sensor is based on a cloverleaf structure with four main washers, as proposed in [10]. They have been fabricated in our cross-type Josephson junction technology [11]. Due to the negligible parasitic capacitance from surroundings of the junction, the total capacitance for the used junction with a size of 0.8 μm × 0.8 μm can be estimated to about $C_{JJ}$ = 40 fF [12]. Thus, even with moderate McCumber parameter large hysteresis-free usable voltage swings $\delta V$ of the SQUIDs can be achieved. Table I lists the main SQUID parameters, which are typical for the presented devices.

The inductance of the niobium thin-film pickup loop $L_p$ is designed to match the input coil inductance $L_{in}$. An outer pickup coil dimension of 29.5 mm × 33.5 mm and a linewidth of 50 μm have been chosen. For an Nb film thickness of 250 nm, Fasthenry simulation results in an inductance of about 171 nH, which is close to the estimated input coil inductance of 180 nH of the current sensor.

### III. Measurement Results

The SQUID magnetometer has been characterized in a dipstick immersed in liquid helium at 4.2 K inside a superconducting and permalloy shield. With a McCumber parameter $\beta_C \approx 0.66$ the maximum usable voltage swing $\delta V$ amounts to about 135 μV.

Noise measurements have been carried out with a commercial directly coupled SQUID electronics [13] providing an input voltage and current noise of $S_V^{1/2}$ = 0.35 nV/Hz$^{1/2}$ and $S_I^{1/2}$ = 6 pA/Hz$^{1/2}$, respectively. In a single stage set-up a white flux noise level of 0.9 μΦ$_0$/Hz$^{1/2}$ has been measured. As the transfer function in the working point corresponds to $V_\Phi \approx 500$ μV/Φ$_0$ this value is still dominated by contributions from the room-temperature electronics $S_V^{1/2}/V_\Phi \approx 0.7$ μΦ$_0$/Hz$^{1/2}$.

In order to exploit the intrinsic noise of the SQUID magnetometer, additional positive feedback has been used [14]. Noise spectra were taken with a HP 3565 spectrum analyzer with a maximum bandwidth of 100 kHz. The contribution of the APF resistor accounts to about 0.2 μΦ$_0$/Hz$^{1/2}$ and was subtracted from the measured noise spectrum in the subsequent discussion. For the SQUID magnetometer a white flux noise of 0.56 μΦ$_0$/Hz$^{1/2}$ has been determined, which still includes noise from room-temperature electronics of about 0.1 μΦ$_0$/Hz$^{1/2}$. This noise corresponds to an energy resolution of 5.5 $h$, with $h$ being Planck's constant. The black line in Fig. 2 shows the according flux noise spectrum of the device.

In addition, noise measurements have been performed in a two-stage readout configuration with a second SQUID as a low noise preamplifier, as described in [15]. Here the magnetometer has been voltage-biased and the amplifier SQUID serves as an ammeter. Nearly identical noise characteristics have been measured in such a way, shown as a blue line in Fig. 2. The small kink at about 1 Hz is probably due to some noise arising from our laboratory.

The effective area $A_{eff}$ of the SQUID magnetometer has been determined in a Helmholtz coil system, as described in [16]. The inverse effective area accounts to about $1/A_{eff}$ = 0.187 nT/Φ$_0$. The according field noise with $S_B^{1/2} = S_\Phi^{1/2}/A_{eff}$ is shown on the right axis in Fig. 2. The magnetic field noise amounts to about 0.1 fT/Hz$^{1/2}$ and 0.9 fT/Hz$^{1/2}$ in the white noise region and at 1 Hz, respectively. The field noise spectrum can be described by

$$S_B^{1/2} = 0.09 \text{ fT/Hz}^{1/2} \cdot [1+(300\text{Hz}/f)^{0.3}+(3\text{Hz}/f)], \quad (1)$$

shown as a red dashed line in Fig. 2.

TABLE I
TYPICAL PARAMETER OF THE SQUID CURRENT SENSORS

| | |
|---|---|
| Chip size [mm] | 2.5 × 2.5 |
| SQUID inductance $L_{SQ}$ [pH] | 180 |
| SQUID critical current $2I_C$ [μA] | 12.4 |
| Shunt resistance $R$ [Ω] | 29 |
| Screening parameter $\beta_L = 2I_C L_{SQ}/\Phi_0$ | 1.12 |
| McCumber parameter $\beta_C = 2\pi I_C R^2 C_{JJ}/\Phi_0$ | 0.66 |
| Input coil inductance $L_{in}$ [nH] | 180 |
| Input coil coupling $1/M_{in}$ [μA/Φ$_0$] | 0.40 |
| Usable voltage swing $\delta V$ [μV] | 135 |

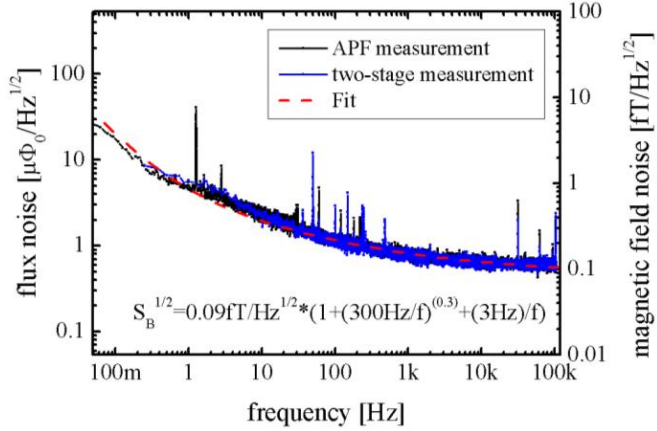

Fig. 2. Flux noise and magnetic field noise spectrum of a SQUID magnetometer composed of a SQUID current sensor and a thin-film pickup coil with dimensions of 29 mm × 33 mm. The black date points belong to the noise measurement with additional positive feedback, the blue to the described two-stage measurement and the red dashed line represents the fit of the noise spectra according to (1).

There seems to be (at least) two independent low-frequency noise sources. Similar observations have been reported as well on other SQUIDs fabricated in this technology [8, 17, 18]. Accordingly, the increase in the frequency range 1 Hz < $f$ < 10 kHz is due to a magnetic signal, whereas below f ≈ 1 Hz the noise is dominated by critical current fluctuations in the Josephson junctions. Revealing the origin of this magnetic noise may further improve the sensor performance.

Nonetheless, the device already provides superior noise performance and may be used in several applications, such as passive geophysical electromagnetic methods, e.g. audiofrequency magnetics or magnetotellurics. We would, however, like to point out that although the SQUID sensor by itself may exhibit such an excellent noise performance, special attention needs to be paid on the surrounding, as e.g. noise arising from the dewar may impair the performance of the SQUID system. In this regard, the presented SQUID magnetometer already offers a tool for the optimization of these components.

## IV. Conclusion

A highly sensitive SQUID magnetometer composed of a niobium thin-film pickup loop and a low noise SQUID current sensor has been presented. The small Josephson junction capacitance results in flux-voltage characteristics with large usable voltage swings and low intrinsic noise of the SQUID. The device exhibits a magnetic field noise of about 0.1 fT/Hz$^{1/2}$ and 0.9 fT/Hz$^{1/2}$ in the white noise region and at 1 Hz, respectively. Investigations on the low-frequency magnetic noise source are ongoing and may further improve the sensor performance.

For application in ULF MRI, the planar pickup loop of the presented magnetometer may be replaced by a gradiometric thin-film version and additional provisions to protect the SQUID from the pre-polarization field may be applied. Our further investigations will focus on the frequency dependent noise behavior of this sensor under applied magnetic fields.